\documentstyle[prl,multicol,aps,psfig]{revtex}
\def\altna{{\mathcal A}_{\alpha}}
\def\altnr{{\mathcal A}_{r}}

\begin{document}
\draft
\widetext
 
\title{Electron correlation in C$_{4N+2}$ carbon rings:
       aromatic vs. dimerized structures}

\author{Tommaso Torelli$^1$ and Lubos Mitas$^2$}
\address{$^1$ Department of Physics,
University of Illinois at Urbana-Champaign, Urbana, Illinois 61801}
\address{$^2$ National Center for Supercomputing Applications,
University of Illinois at Urbana-Champaign, Urbana, Illinois 61801}

\date{\today} 
\maketitle
 
\begin{abstract} 
 The electronic structure of C$_{4N+2}$ carbon rings exhibits competing
 many-body effects of H\"uckel aromaticity, second-order Jahn-Teller and
 Peierls instability at large sizes. This leads to possible ground state
 structures with aromatic, bond angle or bond length alternated geometry.
 Highly accurate quantum Monte Carlo results indicate the existence of
 a crossover between C$_{10}$ and C$_{14}$ from bond angle to bond 
 length alternation. The aromatic isomer is always a transition
 state. The driving mechanism is the second-order Jahn-Teller effect 
 which keeps the gap open at all sizes.
\end{abstract}

\begin{multicols}{2}
\narrowtext

\begin{figure}
\centerline{\psfig{file=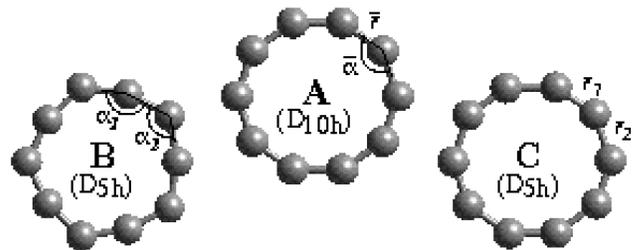,width=3.4in}}
\caption{The most stable isomers of C$_{4N+2}$ rings (shown for
C$_{10}$). The parameters needed to specify the geometries are:
average bond length $\bar{r}=(r_1+r_2)/2$, and bond angle
$\bar{\alpha}=(\alpha_1+\alpha_2)/2$; relative bond length
$\altnr=(r_1-r_2)/\bar{r}$, and bond angle alternation 
$\altna=(\alpha_1-\alpha_2)/\bar{\alpha}$.}
\label{isomers}
\end{figure}

The discovery of carbon fullerenes and nanotubes has opened a 
new materials research area with a vast potential for practical 
applications. Unfortunately, our understanding of the rich 
variety of structural and electronic properties of carbon 
nanostructures is far from satisfactory. For example, experiments~\cite{EXP}
indicate that quasi-one-dimensional structures such as chains and 
rings are among the primary precursors in the formation process 
of fullerenes and nanotubes. However, our insights into their 
properties and behavior are incomplete due to the complicated 
many-body effects involved. In particular, recent studies~\cite{GMR95,KTN99}
have demonstrated a profound impact of the electron correlation 
on stability and other properties of such all-carbon structures.
An important example of such nanostructures is the system of planar
monocyclic carbon rings C$_n$ with $n\!=\!4N\!+\!2$, where $N$ is a
natural number. These closed-shell molecules manifest an intriguing
competition between conjugated aromaticity, second-order
Jahn-Teller and, at large sizes, Peierls instability effects.
Consequently, this leads to different stabilization mechanisms 
that tend to favor one of the following structures: a 
cumulenic ring ({\bf A}), with full D$_{nh}$ symmetry, with 
all bond angles and bond lengths equal; or either of two
distorted ring structures, of lower D$_{{n \over 2}\!h}$ symmetry,
with alternating bond angles ({\bf B}) or bond lengths
({\bf C}). Further structural details are given in Fig.\ \ref{isomers}.
Accurate studies for the smallest sizes (C$_6$ and C$_{10}$)
find isomer {\bf B} to be the most stable.
However, for larger sizes the results from commonly used methods
are contradictory and available experiments~\cite{WKA97} are
unable to clearly specify the lowest energy structures.

In order to identify the most stable isomers and to elucidate
the impact of many-body effects, we carried out an extensive study
of electronic structure and geometries of C$_{4N+2}$ rings of
intermediate sizes up to 22 atoms (with some methods up to 90 atoms).
We employed a number of electronic structure methods including the
highly accurate quantum Monte Carlo (QMC) method which has been proven 
very effective in investigations of C$_{20}$~\cite{GMR95} and larger 
carbon clusters~\cite{KTN99}, as confirmed also by an independent
study by Murphy and Friesner~\cite{MF98}. Our QMC results reveal
that the C$_{4N+2}$ ground state structures have {\em alternated
geometries at all sizes} while cumulenic isomer {\bf A} is
a structural transition state. The results also provide valuable 
insights into the shortcomings of the density functional approaches
such as inaccurate balance between exchange and correlation in
commonly used functionals. In addition, the letter presents a 
first evaluation of interatomic forces in large systems within 
the QMC framework.

According to the H\"uckel rule, the $4N\!+\!2$ stoichiometry implies
the existence of a double conjugated $\pi$-electron system (in-
and out-of-plane). Combined with the ring planarity, this suggests
a strong aromatic stabilization in favor of isomer {\bf A}. Although
the highest occupied and the lowest unoccupied molecular orbitals
(HOMO and LUMO) are separated by a gap of several eV, a double degeneracy 
in the HOMO and LUMO states opens the possibility for a second-order 
Jahn-Teller distortion~\cite{RGP75}, resulting in either cumulenic
{\bf B} or acetylenic {\bf C} structure. Such distortion lowers 
the symmetry and splits the 
degeneracy by a fraction of an eV, with an overall energy gain.
Moreover, as $N\!\to\!\infty$, the picture is complicated further 
by the fact that the system becomes a semimetallic polymer with two 
half-filled $\pi$ bands. As first pointed out by Peierls~\cite{REP55}, 
such a one-dimensional system is intrinsically unstable and undergoes a 
spontaneous distortion which lowers the symmetry. The symmetry 
breaking enables the formation of a gap, in analogy to the elusive 
case of {\it trans}-polyacetylene~\cite{APK89}.

It is very instructive to see how the commonly used computational 
methods deal with such many-body effects. Density functional theory 
(DFT) methods tend to favor a ``homogenized'' electronic
structure with delocalized electrons. In fact, for sizes larger
than C$_{10}$, there is no indication of any stable alternation up
to the largest sizes we have investigated (C$_{90}$). Calculations
performed within the local density approximation (LDA) and
generalized gradient approximations (GGA, with BPW91 functional)
consistently converge to the aromatic structure {\bf A}, in agreement
with other studies~\cite{HLD94}. Only by extrapolation to the
infinite-chain limit, Bylaska, Weare {\it et al.}~\cite{BWK98}
claim to observe a very small, yet stable, bond alternation within 
LDA. A very different picture arises from the Hartree-Fock (HF)
method, which shows a pronounced dimerization for C$_{10}$ and larger.
This agrees with the HF tendency to render structures less
homogeneous in order to increase the impact of exchange effects.
We also verified that using GGA functionals with an admixture of the
exact HF exchange (B3PW91) recovers qualitatively the HF results 
for large sizes ($>$C$_{46}$), as already observed by others~\cite{MEF95}.

Obviously, this problem calls for much more accurate treatments.
High-level post-HF methods, such as multi-configuration
self-consistent field (MCSCF) and coupled cluster (CC), 
indeed provide answers for the smallest ring sizes 
(C$_6$~\cite{HL94} and C$_{10}$~\cite{WB92,MT96}).
In particular, Martin and Taylor~\cite{MT96} have carried out
a detailed CC study demonstrating that both C$_6$ and C$_{10}$
have angle alternated ground state structures, although for
C$_{10}$ the energy of the aromatic isomer {\bf A} is found 
to be extremely close ($1$~kcal/mol). In addition, we have performed 
limited CCSD calculations of C$_{14}$ and have found the dimerized 
isomer to be stable by $\simeq\!6$~kcal/mol. Unfortunately, 
these methods are impractical for larger cases or more 
extensive basis sets~\cite{MEF95}.

The quantum Monte Carlo (QMC) method was used to overcome 
these limitations. This method possesses the unique ability 
to describe the electron correlation explicitly and its
favorable scaling in the number of particles enables us to 
apply it to larger systems~\cite{MSC91}.
In the variational Monte Carlo (VMC) method we construct
an optimized correlated many-body trial wavefunction $\Psi_T$,
given by the product of a linear combination of Slater
determinants and a correlation factor

\begin{equation}
\Psi_T = \sum_n d_n
D_n^\uparrow\{\varphi_\alpha\} D_n^\downarrow\{\varphi_\beta\}
\exp \sum_{I,i<j} u(r_{iI},r_{jI},r_{ij})
\end{equation}

\noindent
where $\varphi$ are one-particle orbitals, $i,j$ denote the 
electrons, $I$ the ions and $r_{iI},r_{jI},r_{ij}$ are the 
corresponding distances. The correlation part, $u$, includes 
two-body (electron-electron) and three-body 
(electron-electron-ion) correlation terms and 
its expansion coefficients are optimized variationally.
Most of the variational bias is subsequently removed by 
the diffusion Monte Carlo (DMC) method, based on the 
action of the projection

\begin{figure}
\centerline{\psfig{file=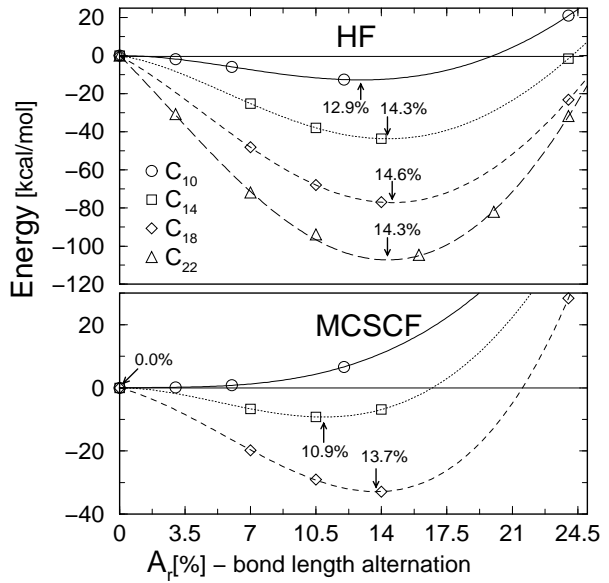,width=3.4in}}
\caption{HF and MCSCF energy as a function of the degree 
of dimerization $\altnr$. Least-squares estimates of the 
positions of the minima are indicated by arrows.}
\label{pes}
\end{figure}

\noindent
operator $\exp(-\tau H)$;
in the limit of $\tau\!\to\!\infty$, this projector 
recovers the lowest eigenstate from an arbitrary trial 
function of the same symmetry and nonzero overlap.
The fermion antisymmetry (or sign) problem is circumvented
by the fixed-node approximation. More details about the 
method are given elsewhere~\cite{MSC91}.

DFT, HF and MCSCF calculations have been carried out using
standard quantum chemistry packages~\cite{GAU98}.
All calculations employed an accurate basis set, consisting 
of $10s11p2d$ Gaussians contracted to $3s3p2d$, and smooth
effective core potentials~\cite{GL98} to replace the 
chemically inert core electrons. 

The geometries of smaller rings with 6 and 10 atoms have already
been established from previous calculations~\cite{HL94,WB92,MT96}.
We have verified that the most reliable published structural 
parameters agree very well (within $\simeq\!0.002$~\AA\ and 
$1^{\circ}$) with our own GGA values.
However, since the dimerized isomer {\bf C} is unstable within DFT,
we followed a different strategy. We began from HF geometries,
which show that the degree of bond length alternation saturates
at $\altnr\!\approx\!14\%$ (Fig.\ \ref{pes}).
In order to correct for the HF bias favoring acetylenic structures,
we performed limited MCSCF calculations (see below) for C$_{10}$,
C$_{14}$, and C$_{18}$. The electron correlation has a profound 
effect on the geometry, to the extent of causing the dimerized 
isomer to be unstable for C$_{10}$, while for C$_{14}$ it decreases 
the dimerization to $\altnr\!\approx\!10\%$. Clearly the limited MCSCF 
for C$_{14}$ and C$_{18}$ provides rather poor geometry 
improvement although one expects a larger correction as more
correlation energy is recovered. In order to verify this and
to estimate the correct degree of dimerization for C$_{14}$, we
carried out the evaluation of the Born-Oppenheimer forces by a 
finite-difference scheme using correlated sampling, in the VMC
method~\cite{HLR94,TM00}. The computation of interatomic forces
is a new development in QMC methodology and, to our knowledge,
this is the first application in this range of system sizes.
We probed the tangential C-C stretching/shortening which leads 
to a change in the degree of dimerization, $\altnr$.
For $\altnr\!=\!7\%$, our force estimate is
$F\!=\!-dE/d\theta\!=\!0.010(2)$~a.u.\ (and a second derivative 
of $H\!=\!0.30(1)$~a.u.), suggesting proximity to the minimum.
Moreover, at $\altnr\!=\!10.5\%$ we find a force of opposite 
sign: $F\!=\!-0.013(3)$~a.u.\ ($H\!=\!0.33(1)$~a.u.).
For C$_{18}$, we have instead performed two QMC single point 
calculations at $\altnr\!=\!7\%,14\%$ and found the first energy
to be lower by $\Delta E\!\simeq\!-12$~kcal/mol. Finally, we 
assumed $\altnr\!=\!7\%$ and $\bar{r}\!=\!1.286$~\AA\ as our 
best estimate for calculations of the acetylenic isomer with $n>10$. 

The crucial ingredient for very accurate QMC calculations is 
a trial function with a small fixed-node error. The quality of 
the one-particle orbitals is of prime importance for decreasing 
such error. While HF or DFT orbitals are commonly used for
construction of Slater determinants, 
our recent projects~\cite{MG97} have demonstrated that 
natural orbitals from limited correlated calculations (e.g.,
MCSCF) lead to more consistent results. Inclusion of 
the electron correlation into the method used to generate 
the orbitals is indeed very relevant for obtaining 
improved fermion nodes, especially for such systems which 
exhibit strong non-dynamical correlation effects~\cite{MEF95,MT96}.
Extensive tests confirmed that orbitals from MCSCF 
(with an active space consisting of 4 occupied and 4 
virtual orbitals) yield the lowest energies and 
so they were used in all of our calculations. In addition,
the inclusion of the most important excited configurations
into $\Psi_T$ (about 20--30 determinants) provided further
significant improvement of the total energies.
In particular, the weights of {\em single} excitations were 
surprisingly large for the alternated geometries and comparable 
to the largest weights of configurations with double excitations.
A more detailed analysis on the multi-reference nature of 
the wavefunction in these systems will be given elsewhere.

\begin{figure}
\centerline{\psfig{file=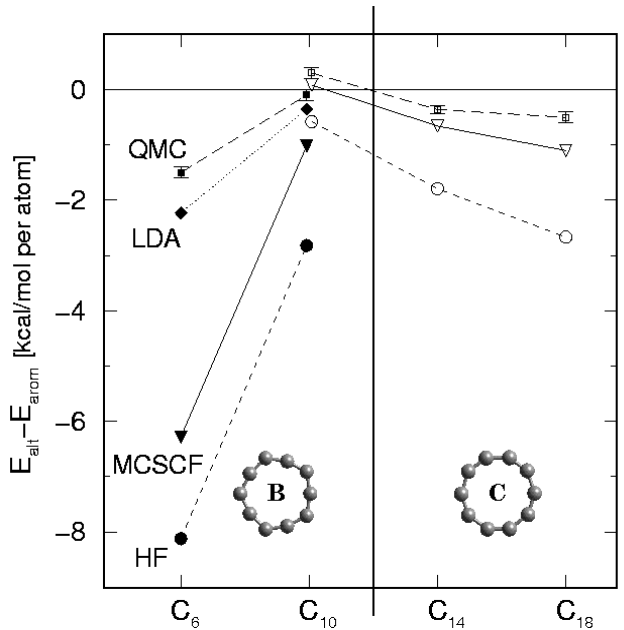,width=3.4in}}
\caption{Relative energies of the angle alternated {\bf B} (filled
symbols) and bond alternated {\bf C} (empty symbols) structures
with reference to the aromatic isomer {\bf A} (see Fig.\ \ref{isomers}).}
\label{rel_en}
\end{figure}

Equipped with such high quality trial functions we have carried
out QMC calculations from  C$_6$ to C$_{18}$. A plot of the energy 
differences, with comparison to other methods, is shown in Fig.\ \ref{rel_en}. 
For the very compact C$_6$ ring, where the overlap between 
$\pi$ in-plane orbitals is large, as observed by Raghavachari
{\it et al.}~\cite{RWP86}, the angle alternated isomer {\bf B}
is the most stable. The aromatic structure {\bf A} is instead a 
transition state leading to angle alternation (B$_{1u}$ mode),
while the dimerized isomer {\bf C} is unstable in all methods.

C$_{10}$ is the case which was studied extensively in the past.
Our DMC results agree with calculations of Martin and Taylor~\cite{MT96}.
We conclude that the angle alternated isomer is still the lowest
energy structure, albeit 

\noindent
extremely close to the cumulenic {\bf A}  
geometry, with an energy difference of $\simeq\!1$~kcal/mol. 
Indeed, the stabilization by aromaticity
is almost as strong as the effect of second-order Jahn-Teller
distortion which is responsible for the alternation pattern.
The aromatic isomer remains a transition state, as it is for 
C$_6$, although in this case the energy surface is extremely
flat. The C$_{10}$ acetylenic isomer appears unstable in DMC, which
implies that our older calculations~\cite{MG97} and also
a more recent all-electron study~\cite{SSL99}, based on a 
single determinant $\Psi_T$ with HF orbitals, were not 
accurate enough.

Our perhaps most interesting results come from the C$_{14}$ 
and C$_{18}$ isomers. The angle alternated structures become
unstable since the in-plane orbital overlap is smaller
due to the increased ring radius. The trends from HF, MCSCF 
and QMC (Fig.\ \ref{rel_en}) are clearly in favor of dimerized
geometries, although there is an indication that the HF bias for
the Jahn-Teller distortion is much reduced as we recover more of 
the correlation energy. Nevertheless, since the fixed-node DMC 
method typically obtains $\approx\!95\%$ of the correlation energy, 
we argue that margins for possible errors are very small.
This is in sharp contrast with the density functional 
results, which indicate only the aromatic isomer {\bf A} to be 
stable. It seems that DFT methods ``overshoot'' the effect
of correlation at the expense of exchange, resulting in a qualitative
disagreement with QMC results.

The data from HF and QMC calculations enable us to analyze the HF
and correlation energy components separately as a function of ring size.
For $n\!\geq\!10$, the HF energy difference between aromatic ({\bf A}) 
and dimerized ({\bf C}, $\altnr=7\%$) isomers can be approximated by

\begin{equation}
E_{HF}^{\bf A} - E_{HF}^{\bf C} 
\simeq 5.14 n - 45.4 \;\; [\textrm{kcal/mol}].
\end{equation}

\noindent
For the correlation energy such an extrapolation is less certain,
as we lack data for very large sizes. The following formula
reproduced our C$_{10}$--C$_{22}$ data within the error bars obtained
(the value for C$_{22}$ was based on an adjusted single determinant 
DMC calculation)

\begin{equation}
E_{corr}^{\bf A} - E_{corr}^{\bf C} 
\simeq - 3.57n + 27.4 \;\; [\textrm{kcal/mol}]
\end{equation}

\noindent
while for larger sizes it is probably somewhat less accurate.
Nevertheless, the significant difference between the coefficients
enables us to recognize the dominance
of the Coulomb and exchange contributions for large $n$. This
means that the competition between the stabilization mechanisms,
which determine the geometry, gap, and key electronic structure
features, is effectively decided {\em at intermediate ring
sizes}. On the basis of these results we suggest that
the ground state isomers are alternated at all sizes.

In conclusion, we present a study of C$_{4N+2}$ carbon rings,
for which highly accurate fixed-node diffusion Monte Carlo
results show that distorted isomers are always preferred with
a stability crossover from bond angle to bond length alternation
between C$_{10}$ and C$_{14}$.
The fully-symmetric aromatic isomer is instead a structural
transition state connecting equivalent alternated geometries
shifted by one site along the ring, although for C$_{10}$ the
angle alternated isomer is below the aromatic one by only
$\simeq\!1$~kcal/mol.
The intermediate size rings (above C$_{10}$) show a clear
trend of dimerization, indicating that the driving
stabilization mechanism is the second-order Jahn-Teller
effect. The corresponding HOMO-LUMO gap persists in large
rings and overshadows thus the onset of the Peierls regime.
The calculations also provide an interesting insight into
the deficiencies of density functional approaches such as 
the imbalance in approximating the exchange and correlation
energy contributions.
The competition of complicated many-body effects, from 
aromatic conjugation to symmetry-breaking, provides an 
extremely sensitive probe of the treatment of electron 
correlation in different theoretical approaches.
Our results demonstrate the potential of quantum Monte Carlo
not only for evaluation of accurate energy differences, but 
also for other properties, such as equilibrium structures, 
which can be obtained by new developments in the QMC methodology.

We are grateful to J. Weare and E. Bylaska for suggesting this 
problem and for discussions.
L.M. would like to thank M. Head-Gordon and R.J. Bartlett for 
stimulating discussions. This research was supported by the DOE 
ASCI initiative at UIUC, by the State of Illinois and by NCSA.
The calculations were carried out on the SPP-2000 Exemplar,
Origin2000, NT Supercluster at NCSA, and the T3E at PSC. We 
also would like to acknowledge R. Reddy at PSC for technical help. 


\end{multicols}

\end{document}